\documentclass[aps,twocolumn,preprintnumbers,nofootinbib,prl,superscriptaddress]{revtex4-2}

\usepackage[utf8]{inputenc}

\usepackage{graphicx,amssymb,amsmath,amsthm,amsfonts}
\usepackage{mathtools}
\usepackage[usenames,dvipsnames,svgnames,table]{xcolor}
\definecolor{darkred}{rgb}{0.5,0,0}
\usepackage{aas_macros}
\usepackage{bm}
\usepackage{dcolumn}
\usepackage{latexsym}
\usepackage{rotating}

\usepackage[normalem]{ulem}

\usepackage{enumerate}
\usepackage{tensor,multirow}
\usepackage{url}
\usepackage[linktocpage]{hyperref}

\newcommand{\gint}{\int d^4x \, \sqrt{-g} \,}

\newcommand{\tder}{\partial_{t}}
\newcommand{\rder}{\partial_{r}}

\usepackage{braket}

\def\be{\begin{equation}}
\def\ee{\end{equation}}
\newcommand{\beq}{\begin{eqnarray}}
\newcommand{\eeq}{\end{eqnarray}}

\newcommand{\GG}{\mathcal G}
\newcommand{\RR}{\mathcal R}

\def\ba{\begin{align}}
\def\ea{\end{align}}

\begin{document}

\title{
What is the fate of Hawking evaporation\\ in gravity theories with higher curvature terms?
}

\author{Fabrizio Corelli}
\email{fabrizio.corelli@uniroma1.it}
\affiliation{Dipartimento di Fisica, ``Sapienza" Universit\`a di Roma \& Sezione INFN Roma1, Piazzale Aldo Moro 
5, 00185, Roma, Italy}

\author{Marina de Amicis}
\affiliation{Dipartimento di Fisica, ``Sapienza" Universit\`a di Roma \& Sezione INFN Roma1, Piazzale Aldo Moro 
5, 00185, Roma, Italy}
\affiliation{Niels Bohr International Academy, Niels Bohr Institute, Blegdamsvej 17, 2100 Copenhagen, Denmark}

\author{Taishi Ikeda}
\affiliation{Dipartimento di Fisica, ``Sapienza" Universit\`a di Roma \& Sezione INFN Roma1, Piazzale Aldo Moro 
5, 00185, Roma, Italy}

\author{Paolo Pani}
\email{paolo.pani@uniroma1.it}
\affiliation{Dipartimento di Fisica, ``Sapienza" Universit\`a di Roma \& Sezione INFN Roma1, Piazzale Aldo Moro 
5, 00185, Roma, Italy}

\begin{abstract}
During the final stages of black hole evaporation, ultraviolet deviations from General Relativity eventually become dramatic, potentially affecting the end-state.
We explore this problem by performing nonlinear simulations of wave packets in Einstein-dilaton-Gauss-Bonnet gravity, the only gravity theory with quadratic curvature terms which can be studied at fully nonperturbative level.
Black holes in this theory have a minimum mass but also a nonvanishing temperature. This poses a puzzle concerning the final fate of Hawking evaporation in the presence of high-curvature nonperturbative effects.
By simulating the mass loss induced by evaporation at the classical level using an auxiliary phantom field, we study the nonlinear evolution of black holes past the minimum mass. We observe a runaway shrink of the horizon (a nonperturbative effect forbidden in General Relativity) which eventually unveils a high-curvature elliptic region.
While this might hint to the formation of a naked singularity (and hence to a violation of the weak cosmic censorship) or of a pathological spacetime region, a different
numerical formulation of the initial-value problem in this theory might be required to rule out other possibilities, including the transition from the critical black hole to a stable horizonless remnant.
Our study is relevant in the context of the information-loss paradox, dark-matter remnants, and for constraints on microscopic primordial black holes. 

\end{abstract}

\maketitle

\noindent{{\bf{\em Introduction.}}}
It is widely believed that some of the deepest theoretical problems of General Relativity~(GR) --~such as its non-renormalizability and the unavoidable formation of singularities~\cite{HawkingEllis,1969NCimR...1..252P}~-- can be resolved by some ultraviolet completion of which GR is the low-energy approximation. 
Ultraviolet deviations from GR are expected to become important at the Planck scale, although the huge hierarchy between the latter and the curvature regimes probed so far does not exclude that the fundamental ultraviolet energy scale is smaller.
If the beyond-GR scale is (nearly) Planckian it would be virtually impossible to probe it in the foreseeable future (but see~\cite{Addazi:2021xuf}). Nevertheless, there is a well-established semiclassical process that can  in principle provide a portal to such energy scale, namely black-hole~(BH) Hawking evaporation~\cite{Hawking:1975vcx}. The fact that a static BH with mass $M_{\rm BH}$ emits radiation nearly as a blackbody with temperature $T_H= \frac{\hbar c^3}{8\pi G k_B M_{\rm BH}}$ (where $c$, $G$, $\hbar$ and $k_B$ are all fundamental constants that we set to unit henceforth) is at the root of the infamous BH information loss paradox~\cite{Hawking:1975vcx,Mathur:2009hf,Polchinski:2016hrw}, and suggests that BH evaporation (and the physics at the high-curvature scales where evaporation is relevant) is a portal connecting gravity, quantum theory, relativity, and thermodynamics~\cite{Bekenstein,Hawking:1976de}.

During Hawking evaporation, the BH gradually shrinks probing near-horizon regions of ever growing curvature. It is therefore inevitable that ultraviolet deviations from GR eventually become dramatic during this process.
Studying the evolution of BH Hawking evaporation beyond GR within an effective field theory approach is of limited interest, because deviations from the standard GR picture ought to be perturbative, while the most interesting effects are expected to occur in a nonperturbative regime. However, studying the nonlinear dynamics of gravity in the high-curvature regime beyond GR at the nonperturbative level is very challenging~\cite{Foucart:2022iwu}.

In this Letter and in a companion paper~\cite{companion}, we report on a substantial step forward in this direction. By making use of recent progress~\cite{Ripley:2019aqj,Ripley:2019hxt,Ripley:2019irj,East:2020hgw,East:2021bqk,Kuan:2021lol,Kuan:2021yih} in evolving dynamical spacetimes in nonperturbative quadratic gravity, we shall investigate the fate of mass evaporation in the presence of nonperturbative high-curvature deviations from GR.

Our starting point is Einstein-scalar-Gauss-Bonnet gravity~\cite{Kanti:1995vq}, 
\begin{align}
	S &= \frac{1}{16\pi}\gint \biggl\{ \RR - \bigl(\nabla \phi\bigr)^2 + 2 F[\phi] \GG\biggr\} + S_m\,,
	\label{eq:Action}
\end{align}
where $\RR$ is the scalar curvature, $\phi$ is the dilatonic field, $\GG = \frac{1}{4} \epsilon^{\mu\nu\alpha\beta} \epsilon_{\rho\sigma\lambda\omega} \tensor{R}{^{\rho\sigma}_{\mu\nu}} \tensor{R}{^{\lambda\omega}_{\alpha\beta}}$ is the Gauss-Bonnet invariant, $F[\phi]\propto \lambda$ is a function that depends on the fundamental coupling constant $\lambda$, and $S_m$ is the matter action. Curvature-squared corrections to GR are ubiquitous in both bosonic and heterotic string theories~\cite{Gross:1986mw}, but~\eqref{eq:Action} is the only quadratic gravity theory with second-order field equations, thus avoiding the Ostrograski's instability~\cite{Woodard:2006nt} even when considered at the nonperturbative level.

Studying BH evaporation in this theory is particularly interesting also for another reason: due to nonperturbative effects, in this theory BHs only exist above a minimum-mass solution~\cite{Kanti:1995vq,Torii:1996yi,Alexeev:1996vs,Pani:2009wy}, $M_{\rm BH}\geq M_{\rm crit}\propto\sqrt{\lambda}$, which depends on the coupling function. 
This is a striking difference with respect to GR, where the BH mass is an unconstrained free parameter.
Remarkably, the Hawking temperature~\cite{Torii:1996yi,Alexeev:1996vs} and the graybody factor of the minimum mass solutions are finite~\cite{MarinaThesis,companion}. This implies that in this theory a BH loses its mass at the rate~\cite{gibbons1993action}
\begin{align}
 \frac{dM_{\rm BH}}{dt}=-\frac{1}{2\pi}\sum_{l,m}\int d\omega\frac{\omega G_{lm}(\omega)}{e^{\omega/T_H}\pm 1}\,,
\end{align}
where $G_{lm}(\omega)$ is the graybody factor for the emission of a mode with frequency $\omega$ and $(l,m)$ angular dependence, and the $+/-$ sign refers to the emission of fermions/bosons. This mass loss occurs also near and at the minimum mass, which is inevitably reached during the evaporation, no matter how small the fundamental coupling $\lambda$ is.
A natural question then arises: \emph{what is the final fate of Hawking evaporation in this regime?}~\cite{Torii:1996yi,Alexeyev:2002tg}.

For the first time, we have explored this problem by performing extensive nonlinear evolutions of BHs in the theory~\eqref{eq:Action} in spherical symmetry. Here we report the salient features, a more detailed analysis is presented in~\cite{companion}.

\noindent{{\bf{\em Numerical procedure.}}}
Hawking evaporation makes the BH mass decrease in time, a process that is generically not allowed for reasonable classical interactions (excluding superradiance~\cite{Brito:2015oca} which anyway does not occur in our setup). 
To mimic mass loss due to Hawking evaporation at the classical level, we introduce a minimally-coupled, massless ``phantom'' scalar field $\xi$ whose kinetic term has the wrong sign relative to our conventions, thus violating the energy conditions. The matter action in~\eqref{eq:Action} then reads
\begin{equation}
	S_m = \frac{1}{16\pi} \gint \bigl( \nabla \xi \bigr)^2\,.
	\label{eq:MatterAction}
\end{equation}
We simulate the spherical collapse of wave packets of $\phi$ and $\xi$ on an initially static dilatonic BH in this theory using horizon penetrating coordinates to monitor also the BH interior. For this purpose we adopt a setup similar to the one proposed in Ref.~\cite{Ripley:2019aqj} for evolving BH spacetimes in shift-symmetric GB gravity using Painlev\'e-Gullstrand-like coordinates. We write the line element as
\begin{equation}
	ds^2 = -\alpha^2 dt^2 + (R'(r) dr + \alpha \zeta \, dt)^2 + R(r)^2 d\Omega^2,
	\label{eq:PGLineElementRT}
\end{equation}
where the lapse $\alpha$ and the function $\zeta$ depend on $(r, t)$, and $R(r)$ is the areal radius, defined in order to achieve better resolution in high-curvature regions while keeping a uniform grid for the coordinate radius $r$~\cite{companion}.
After introducing the auxiliary variables $Q = \rder \phi$ and $\Theta = \rder \xi$, 
and the conjugate momenta of the scalar fields, $P = \frac{1}{\alpha} \tder \phi - \frac{\zeta Q}{R'(r)}$ and $\Pi = \frac{1}{\alpha} \tder \xi - \frac{\zeta \Theta}{R'(r)}$,
we obtain a nonlinear system of $7$ evolution equations for $\phi$, $Q$, $P$, $\xi$, $\Theta$, $\Pi$, $\zeta$, and $2$ constraints for $\alpha$ and $\zeta$. 
We evolve this system using the method of lines. We compute the spatial derivatives with the fourth-order accurate centered finite-difference method, and perform the time integration with the fourth-order accurate Runge-Kutta method. At each integration step we update the profile of $\alpha$ by integrating its constraint with a combination of the Simpson's rules. We add a fifth order Kreiss-Oliger term in the right-hand side of the evolution equations~\cite{Babiuc:2007vr} to stabilize the algorithm against high-frequency modes arising from the inner region, and using a weighting function that restricts the dissipative action to the central region.

A noteworthy aspect of this theory is that, at least in this formulation, there might appear regions in which the system is not hyperbolic~\cite{Ripley:2019hxt,Ripley:2019irj,Ripley:2019aqj}. We therefore implement an excision strategy to remove the elliptic regions, which are initially confined within the BH horizon. In particular, at each time step we set the excision boundary as the outermost radius where the characteristic velocities have an imaginary part~\cite{Ripley:2019irj,companion}. We integrate the system only in the hyperbolic region, and thus, by construction, the excised region can never shrink during the evolution.
We stop the simulations if the excision boundary crosses the apparent horizon, since in this case the elliptic region might be causally connected to the BH exterior and the evolution can be pathological~\cite{eventhorizon}.

We prescribe initial data by numerically solving the field equations at $t=0$. For any coupling function such that $F'[\phi=0]\neq0$, BH solutions in this theory are different from their GR counterpart and feature a secondary~\cite{Berti:2015itd,Herdeiro:2015waa} scalar hair. We first construct a numerical solution for the dilatonic BH by fixing the horizon radius and coupling constant and implementing a shooting procedure~\cite{Kanti:1995vq,Pani:2009wy}. 
We fix the units such that the horizon areal radius of the initial BH is $R_h(t = 0) = 2$, which corresponds to setting the initial BH mass to unity in the GR limit. 
Then, we add some ingoing Gaussian wave packets for the dilaton field, with amplitude
\begin{equation}
 \delta \phi = \frac{A_{0, \phi}}{R} \exp\left[-\frac{(R - R_{0, \phi})^2}{\sigma_\phi^2}\right]\,,
\end{equation}
and similarly for the phantom field $\delta \xi$, where $A_{0, \phi}$, $R_{0, \phi}$, and $\sigma_\phi$ (and their phantom-field analog $A_{0, \xi}$, $R_{0, \xi}$, and $\sigma_\xi$) are free parameters.
Finally, we compute the initial profiles of $\alpha$ and $\zeta$ through a numerical integration of the two constraints.
More details on the evolution scheme and code testing are provided in~\cite{companion}.

\begin{figure}[ht]
	\centering
	\includegraphics[width = \columnwidth]{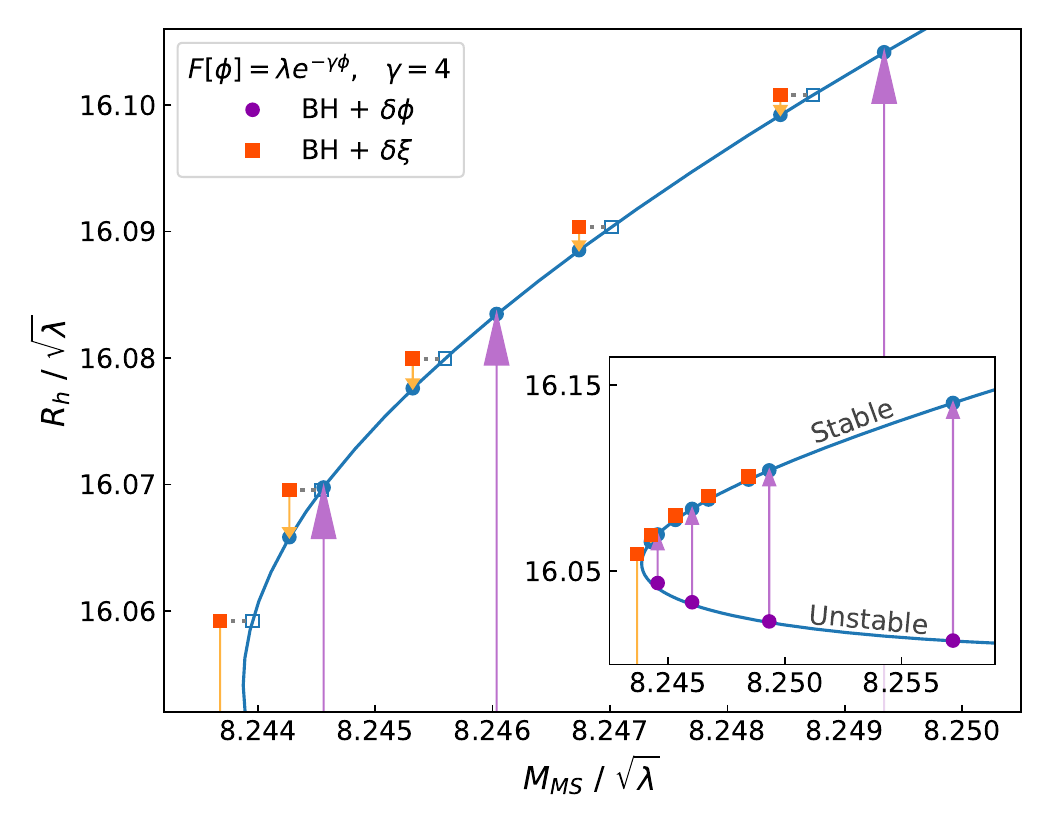}
	\caption{
		Evolution of dilatonic BHs in EdGB gravity in the $R_h-M_{\rm MS}$ plane, where $R_h$ is the horizon areal radius and $M_{\rm MS}$ is the Misner-Sharp mass at infinity. The blue solid curve represents a family of static BHs which features a minimum mass separating two branches of solutions. Purple arrows correspond to the dynamics of a dilaton perturbation on initially static BHs in the lower (unstable) branch, while yellow arrows correspond to the dynamics of a phantom field on BH configurations in the upper (stable) branch. 
		}
	\label{fig:SummaryPlot}
\end{figure}

\noindent{{\bf{\em Results.}}}
For concreteness, we focus on the relevant case of dilatonic BHs in Einstein-dilaton-Gauss-Bonnet~(EdGB) gravity~\cite{Kanti:1995vq} with $F[\phi]=\lambda e^{-\gamma\phi}$, setting $\gamma=4$ (different values of $\gamma\gtrsim1$ give the same qualitative picture~\cite{companion}).
Because of nonperturbative effects, in this theory there is a critical value of the mass below which no BH solutions exist.
The minimum-mass BH divides two branches of solutions with the same mass and different areal radii, see inset of Fig.~\ref{fig:SummaryPlot}. Similarly to the standard case of compact stars in GR~\cite{ShapiroTeu},
one branch is linearly stable and the other (with lower radii) is linearly unstable~\cite{Torii:1998gm}. 
While inside any of these BH solutions there is a curvature singularity~\cite{Alexeev:1996vs}, for the solution at the end of the unstable branch such singularity coincides with the horizon and becomes naked (see, e.g.,~\cite{Sotiriou:2013qea,Sotiriou:2014pfa}). However, the singular solution does \emph{not} coincide with the minimum-mass solution~\cite{Torii:1996yi, Guo:2008hf, MarinaThesis, Blazquez-Salcedo:2017txk}. Therefore, the latter is regular at and outside the horizon, just as in the GR case, and can be treated (semi)classically if its lengthscale is sufficiently far from being Planckian.
The stability analysis just reported has been derived within linear-perturation theory~\cite{Torii:1998gm}. Within our framework we can now explore the full nonlinear dynamics on both the stable and unstable branches.

We first start by simulating wave packets of the dilaton on initially static BHs in the lower branch. We consider the couplings $\lambda= \{1.554, 1.556, 1.558, 1.56\}\times 10^{-2}$, which correspond to an initial BH mass such that $M_{\rm BH}/M_{\rm crit}-1\approx\{0.6, 2.4, 6.4, 15.9\}\times 10^{-4}$, where $M_{\rm crit}\approx 8.244\sqrt{\lambda}$ for this specific coupling function. The initial dilaton perturbation $\delta \phi$ is defined by
\begin{equation}
	A_{0, \phi} = 0.01  , \qquad R_{0, \phi} = 15 , \qquad \sigma_\phi = 2.5 .
	\label{eq:DilatonPerturbationParametersLowerBranch}
\end{equation}
We set $A_{0, \xi}=0$ so that the phantom field vanishes everywhere and the theory reduces to the standard EdGB gravity. 
We place the outer boundary at $R_\infty = 2850$, and set the final time of integration to $T = 2800$. The grid step is $\Delta r = 0.02$ with a Courant-Friedrichs-Lewy factor ${\rm CFL} = 0.025$.

Figure~\ref{fig:SummaryPlot} shows, with purple arrows, the evolution in the $R_h - M_{\rm MS}$ plane (where $M_{\rm MS}$ is the Misner-Sharp mass function computed at spatial infinity~\cite{Ripley:2019aqj}) for this set of simulations. The purple points represent the initial configurations, which are close to the static BH solutions (blue solid curve). Although not visible on the scale of the plot, at the beginning of the simulations the total mass is slightly larger than the corresponding isolated BH mass, since the perturbation $\delta \phi$ adds a positive contribution to $M_{\rm MS}$. As clear from Fig.~\ref{fig:SummaryPlot}, BHs in the lower branch are unstable and migrate towards a final stable configuration in the upper branch.
The blue points in Fig.~\ref{fig:SummaryPlot} represent the end-states of the numerical integration; the fact that they lie on top of the blue curve is a consistency check for our simulations.
These results numerically prove the instability of the lower branch and the stability of the upper branch at the fully nonlinear level. 

\begin{figure}
	\centering
	\includegraphics[width = \columnwidth]{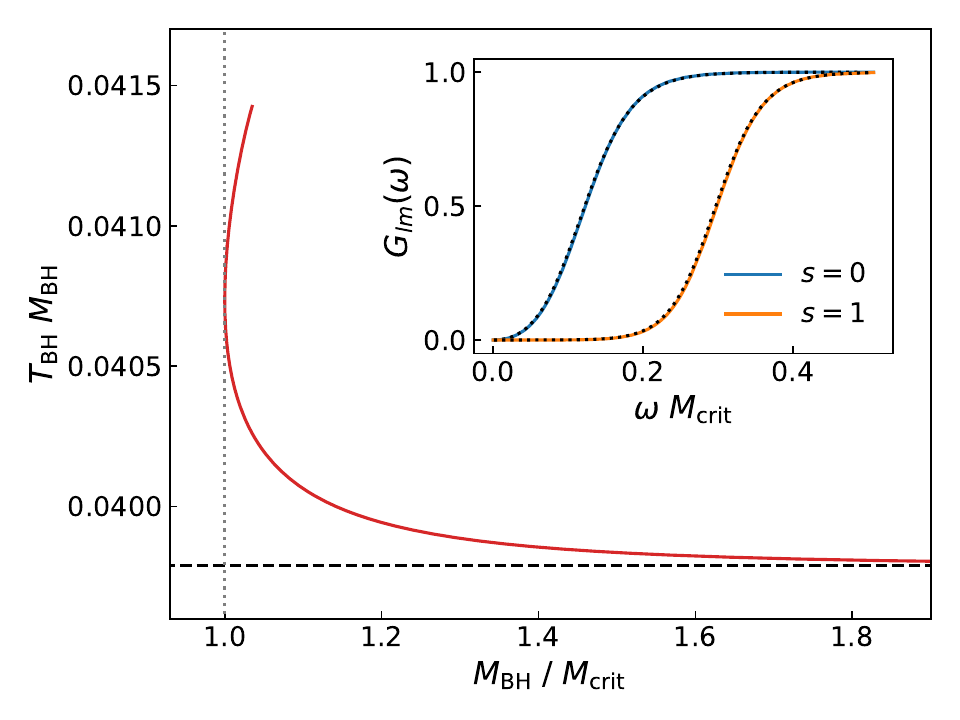}
	\caption{Hawking temperature of a dilatonic BH in EdGB gravity as a function of its mass. The inset shows the graybody factor of the minimum-mass BH solution for the emission of massless scalar particles (blue) and photons (orange) in their lowest angular modes ($l=0,1$, respectively). The dotted lines correspond to the graybody factor of a Schwarzschild BH with the same mass. See~\cite{companion} for details. 
	}
	\label{fig:TemperatureStatic}
\end{figure}

We now move to investigate the dynamics of dilatonic BHs under a mass loss due to absorption of the phantom field. Figure~\ref{fig:TemperatureStatic} shows the Hawking temperature and graybody factor of a dilatonic BH in EdGB gravity, computed in detail in Ref.~\cite{companion} (see also~\cite{Alexeev:1996vs,Konoplya:2019hml}). Overall, they are quantitatively very similar to their GR counterpart even for large coupling constants, suggesting that the rate of mass loss due to Hawking emission in EdGB gravity is similar to the GR case. In fact, near the minimum mass solution mass loss occurs $\approx 10\%$ faster than in GR.

In order to mimic Hawking evaporation we simulate wave packets of the phantom field on initially static dilatonic BHs in the upper branch, in the absence of perturbations of the dilaton ($A_{0, \phi} = 0$). We set the parameters of $\delta \xi$ to 
\begin{equation}
	A_{0, \xi} = 0.01, \qquad R_{0, \xi} = 15 , \qquad \sigma_\xi = 2.5,
	\label{eq:FixedAVaryingLParameters}
\end{equation}
and we consider 5 different coupling constants: $\lambda = \{1.543, 1.545, 1.547, 1.549, 1.551\}\times 10^{-2}$. In the latter case the total Misner-Sharp mass of the spacetime is below the critical one, so after the absorption of the phantom wave packet the BH becomes subcritical. We use the same grid resolution as in the previous simulations.
Other choices of the initial data and setups, including one with an ingoing phantom perturbation and an outgoing standard field perturbation both starting near the horizon (which would more closely mimic the production of a Hawking quantum pair) give the same dynamics as presented below, see~\cite{companion} for details.

Note that the role of the phantom field is solely to dynamically reduce the BH mass, mimicking the effect of Hawking evaporation at the classical level. Furthermore, its evolution is not pathological in spherical symmetry, since absence of gravitational-wave emission prevents runaway instability from free phantom fields. We have checked this explicitly in the GR limit~\cite{companion}. Indeed, as shown by the yellow arrows in Fig.~\ref{fig:SummaryPlot}, when the initial BH is sufficiently far from the critical mass, the initial phantom perturbation is simply absorbed by the horizon and the BH reaches a stable stationary configuration with slightly smaller mass, as expected. The red squares in Fig.~\ref{fig:SummaryPlot} represent the initial configurations, which have a slightly smaller mass than in the isolated BH case (empty squares, connected to the red ones by horizontal dotted lines) since in this case the phantom perturbation $\delta\xi$ gives a negative contribution to $M_{\rm MS}$. Overall, we observe that when the initial configuration is supercritical ($M_{\rm MS}(t=0)>M_{\rm crit}$) the system reaches a stable configuration in the upper branch upon absorbing the phantom field. On the other hand, for subcritical masses ($M_{\rm MS}(t=0)<M_{\rm crit}$), after the BH has initially absorbed the phantom perturbation, on a much longer time scale the apparent horizon starts shrinking and the excision boundary starts expanding due to the intrinsic (classical) dynamics of the theory. The evolution of these two radii becomes faster as the simulations proceeds, until they eventually cross each other and the simulation is stopped when an elliptic region emerges from the horizon.

\begin{figure}[th]
	\centering
    \includegraphics[width = 1.1 \columnwidth]{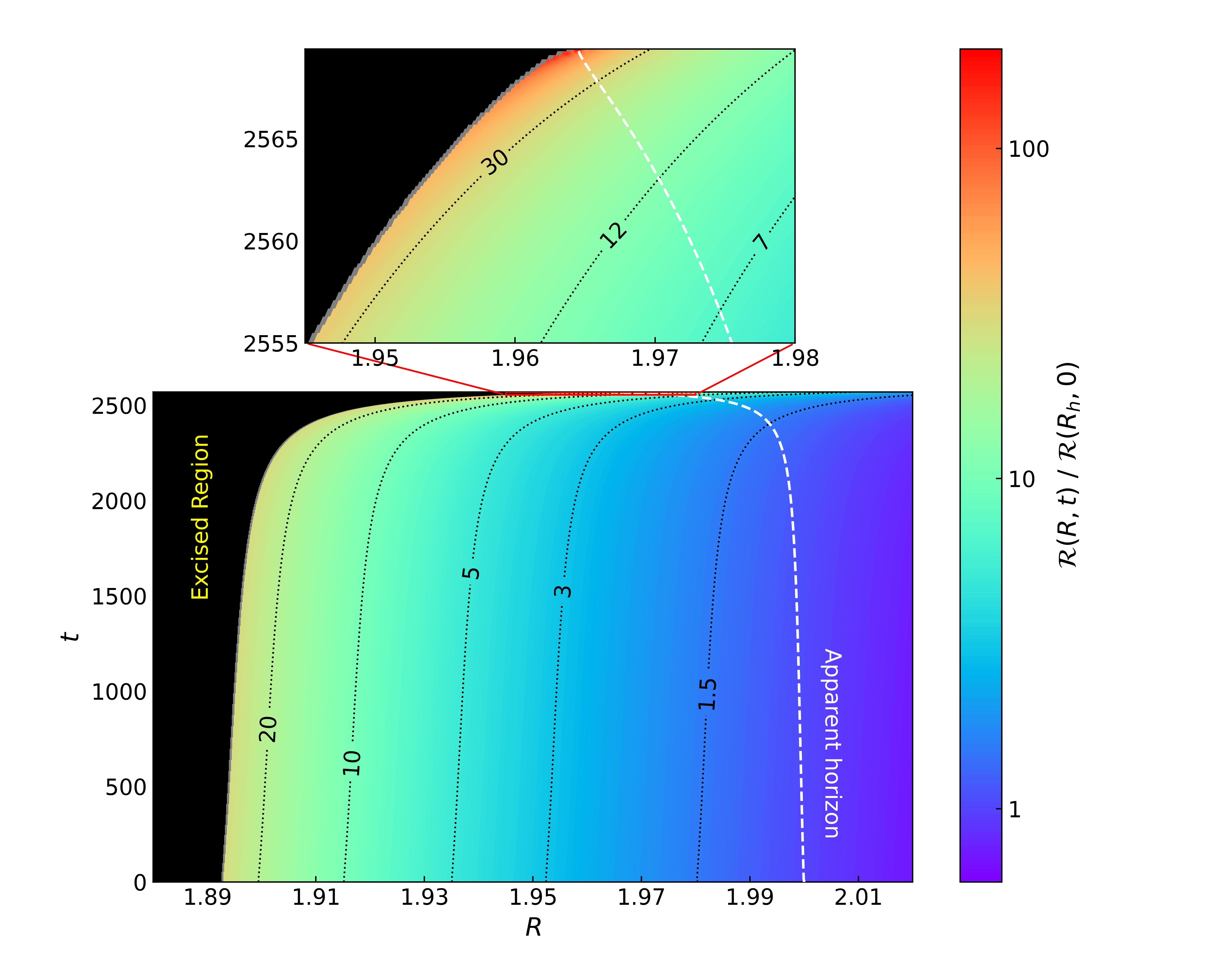}
	\caption{The Ricci curvature scalar $\RR$ as a function of spacetime for a BH absorbing a negative-energy wave packet near the minimum mass solution in EdGB gravity.
	The black region denotes the excision in our simulation, the thin gray area denotes the few grid points near the excision where we avoid computing the curvature, whereas the white dashed curve shows the shrink of the apparent horizon.
	The inset shows the final stages of the simulation when the apparent horizon crosses the excision, unveiling a high-curvature region where the system of partial differential equations becomes elliptic.
    }
	\label{fig:RicciColorPlot}
\end{figure}

\noindent{{\bf{\em Naked singularity formation in EdGB gravity?}}} 
We repeated the simulation in the $M_{\rm MS}(t=0)<M_{\rm crit}$ case increasing the resolution ($\Delta r = 0.0025$), in order to perform a more accurate analysis. In particular, we wish to understand if the expansion of the elliptic region is also related to an increase of the location of the curvature singularity therein, and if the rapid decrease of the apparent horizon radius is leading to an exposure of the singularity and a violation of the weak cosmic censorship conjecture~\cite{1969NCimR...1..252P,Wald:1997wa}. For this purpose we computed the profile of the Ricci scalar $\RR$ at each time step, see Fig.~\ref{fig:RicciColorPlot}. The black area represents the excised region, and the white dashed curve is the apparent horizon. The thin gray area contains the first three grid points outside the excision radius where, conservatively, we did not compute $\RR$ due to the change in the differentiation and dissipation schemes.

By the time the horizon almost crosses the excision, the Ricci scalar at the apparent horizon has grown by almost two orders of magnitude compared to its initial value. In this regime the curvature converges well for different resolutions until $t=2569.0$~\cite{companion}. This suggests that a large curvature region located just across the excision is emerging out of the apparent horizon.
We have indication that the curvature loses convergence only in the very last time steps before the simulation stops (approximately at $t=2569.6$).
Furthermore, by tracing null rays backwards in time during the last stages of the simulations~\cite{companion}, we have checked that the \emph{event} horizon follows the same behavior of the apparent horizon, confirming the general picture.
Unfortunately, the curvature singularity (which exists also at $t=0$ inside the BH, just as in the GR case) is always located inside the excision boundary, so with this formalism we cannot access the region where $\RR$ actually diverges. Nevertheless, the level curves in Fig.~\ref{fig:RicciColorPlot} show that the region of high curvature expands following the behavior of the excision boundary, in particular the high-curvature region quickly moves outwards in the last stages of the simulations. This suggests that the radius of the curvature singularity (that for the minimum-mass solution is initially already close to the outer boundary of the elliptic region~\cite{companion}) expands tracking the evolution of elliptic boundary, thus supporting the hypothesis of the formation of a naked singularity.

\noindent{{\bf{\em Discussion.}}}
Whether or not a naked singularity forms as the outcome of mass loss of a minimum-mass BH in EdGB gravity, we can conclude that a high-curvature elliptic region does emerge out of the apperent horizon, which is a potential major drawback for the theory. However, the appearance of the naked elliptic region depends on the gauge choice~\cite{Ripley:2019hxt,Ripley:2019irj,Bernard:2019fjb,Kovacs:2020pns,Kovacs:2020ywu} and we cannot exclude that a different evolution scheme could cure this pathology~\cite{companion} (see Ref.~\cite{R:2022hlf} for a gauge-independent characterization of elliptic regions in spherical symmetry).
In the companion paper~\cite{companion}, we show that the minimum-mass BH solution actually co-exists in the phase space of the theory with a regular wormhole~\cite{Kanti:2011jz}. This intriguing feature might suggest the possibility of a transition from this critical BH solution to a regular horizonless remnant, which cannot evaporate any further (see also~\cite{Alexeyev:2002tg,Alexeyev_model} for a model in which Hawking evaporation is halted). This would require a change of topology which might be connected to the limitations encountered by our evolution scheme.

Thus, while the question posed in the title remains unanswered, we argue that investigations in this direction would be very valuable. Given the absence of BH solutions with $M_{\rm BH}<M_{\rm crit}$, Hawking evaporation in EdGB gravity is bound to either violate the weak cosmic censorship (implying an inevitable breakdown of the theory and the need of a full quantum gravity completion) or produce (potentially classical) horizonless remnants. Both options are extremely intriguing and deserve further investigation. 
In particular, microscopic horizonless remnants evade all the constraints on light BHs~\cite{Carr:2020gox} arising from Hawking evaporation and could form the entirety of the dark matter. Furthermore, the expectation that primordial BHs formed in the early universe with masses below $M_{\rm BH}\sim 10^{15}\,{\rm g}$ should be completely evaporated by the present epoch and cannot therefore contribute to the dark matter is based on the assumption that GR is valid all the way down to full evaporation, way beyond the curvature scales where the theory has been tested. In the final stage of the evaporation, the ultraviolet terms explored here become dominant. 
Our setup might provide a concrete first-principle model to establish whether the information allegedly lost~\cite{Hawking:1975vcx,Mathur:2009hf,Polchinski:2016hrw} at the end of the evaporation can be stored in stable remnants~\cite{Chen:2014jwq}.

Finally, notice that the EdGB length scale $\sqrt{\lambda}$ might be much larger than the Planck length. Therefore, at variance with GR, the puzzle of Hawking evaporation in this theory might be resolvable without invoking full-fledged quantum-gravity effects. In this context, an important point that we intend to explore in the future is the impact of higher-order terms in the action, which are natural in the ultraviolet regime, on the nonperturbative effects and solutions discussed here and in~\cite{companion}.

\noindent{{\bf{\em Acknowledgments.}}}
We are grateful to Vitor Cardoso, Daniela Doneva, Will East, Leonardo Gualtieri, Luis Lehner, Frans Pretorius, Justin Ripley, and Helvi Witek for useful comments on the draft.
We acknowledge financial support provided under the European Union's H2020 ERC, Starting Grant agreement no.~DarkGRA--757480. 
Computations were performed at Sapienza University of Rome on the Vera cluster of the Amaldi Research Center.
This project has received funding from the European Union’s Horizon 2020 research and innovation programme under the Marie Skłodowska-Curie grant agreement No 101007855.
We also acknowledge support under the MIUR PRIN and FARE programmes (GW-NEXT, 
CUP:~B84I20000100001,  2020KR4KN2), and from the Amaldi Research Center funded by the MIUR program "Dipartimento di Eccellenza" (CUP:~B81I18001170001). This work is partially supported by the PRIN Grant 2020KR4KN2 ``String Theory as a bridge between Gauge Theories and Quantum Gravity''.
\bibliographystyle{apsrev4}
\bibliography{References}

\begin{thebibliography}{47}
\bibitem [{eve()}]{eventhorizon}%
  \BibitemOpen
  \href@noop {} { }\bibinfo {note} {Note that the appearance of an elliptic
  region outside the apparent horizon is not necessarly pathological since, in
  dynamical situations, it cannot be excluded that an elliptic region outside
  the apparent horizon remains confined within the event horizon.}\BibitemShut
  {Stop}%
\end{thebibliography}

\end{document}